\documentclass[12pt]{article}
\usepackage{revsymb}
\usepackage{amsmath,amsfonts,amssymb}
\newcommand{\diag}{\operatorname{diag}}

\title{Optics of Rotating Systems\thanks{Dedicated to the memory of John A. Wheeler (1911-2008).}}

\author{Bahram Mashhoon\\Department of Physics and Astronomy\\University of Missouri\\
Columbia, Missouri 65211\\USA}

\begin{document}

\maketitle

\begin{abstract}
Electrodynamics of rotating systems is expected to exhibit novel nonlocal
features that come about when acceleration-induced nonlocality is
introduced into the special relativity theory in conformity with the
Bohr-Rosenfeld principle. The implications of nonlocality for the
amplitude and frequency of electromagnetic radiation received by uniformly
rotating observers are investigated.
\end{abstract}

\section{Introduction}

The modern theories of special and general relativity have their origin in the problems associated with the way electromagnetic waves appear to observers in motion; in particular, the aberration of starlight provided the initial quandary \cite{1,2}. The purpose of this paper is to argue that certain difficulties still remain and need urgent experimental attention. The proposed theoretical resolution of these new problems calls for the development of nonlocal theories, where nonlocality is induced by the acceleration of the observer~\cite{3}. We use the term ``observer" in an extended sense; the observer could be a sentient being or a measuring device. From a practical standpoint, the main physical assumptions involved in the actual design of electrical devices have been discussed in Ref.~\cite{4}. Furthermore, excellent reviews of the optics of rotating systems are available~\cite{5,6}; however, this paper is about certain new aspects that are specifically due to rotation-induced nonlocality. In previous work, the consequences of nonlocal electrodynamics of rotating
systems have been worked out in detail when radiation is perpendicularly
incident on the orbit of the rotating observer; therefore, the present
treatment is devoted to the general case of oblique incidence. Moreover, the
new effects are discussed here within the more general framework of nonlocal
special relativity.

The standard special-relativistic theory of reception of electromagnetic radiation by a general accelerated observer in Minkowski spacetime is presented in section~\ref{sec:2}. This theory has to be revised, as necessitated by recent theoretical and observational developments, in order to take due account of the photon helicity-rotation coupling. The revised theory is presented in section~\ref{sec:3} and its observational consequences are briefly described. The locality postulate, upon which the theories of sections~\ref{sec:2} and \ref{sec:3} are based, implies that an observer can determine the electromagnetic field \textit{instantaneously}. In fact, the pointwise determination of physical quantities is the gist of the locality postulate. This is contrary to the Bohr-Rosenfeld principle described in section~\ref{sec:4}. According to this principle, only spacetime averages of the electromagnetic field components are physically meaningful. While the Bohr-Rosenfeld requirement is essentially innocuous and inconsequential for inertial observers, this is \textit{not} the case for accelerated observers. Indeed, taking the Bohr-Rosenfeld viewpoint seriously leads to nonlocal special relativity. To explain clearly the theoretical necessity of acceleration-induced nonlocality, we emphasize conceptual issues in sections~\ref{sec:2}-\ref{sec:4}, while keeping the formalism to a minimum. In section~\ref{sec:5}, we discuss uniformly rotating observers in order to illustrate the observational consequences of the nonlocal theory. Finally, section~\ref{sec:6} contains a brief discussion of our results. Throughout, the Minkowski metric tensor is $\diag (-1,1,1,1)$ in our convention and, unless otherwise specified, units are chosen such that $c=1$.

\section{Locality: Standard Approach}\label{sec:2}

In the special theory of relativity, Lorentz invariance is extended to accelerated observers by postulating that such an observer is pointwise inertial. Thus the accelerated observer is in effect replaced by a continuous infinity of otherwise identical momentarily comoving inertial observers. This hypothesis of locality plays a basic role in special relativity as well as in the transition from special to general relativity via Einstein's principle of equivalence. The origin and limitations of the locality postulate have been discussed in Ref.~\cite{7}.

Lorentz invariance is a basic symmetry that involves ideal inertial observers. It is crucial to recognize that all actual observers are accelerated. To relate observation with theory, it is necessary to establish a connection between actual accelerated observers and ideal inertial observers. This connection is postulated to be \textit{nonlocal} in the new theory~\cite{3}. The pointwise \textit{local} equivalence postulated in the standard treatment originates from Newtonian mechanics---same positions and velocities at the same time---and is the simplest approximation, just as a straight inertial worldline tangent to a curved timelike worldline at an event is the simplest Frenet approximation to the curved path at the event.

In the standard approach, an accelerated observer is thus assumed to be instantaneously inertial and at rest in an inertial frame of reference that is boosted with its instantaneous velocity with respect to the background global inertial frame. The inhomogeneous Lorentz transformation $x^{'\mu} =C^\mu +L^\mu_{\;\;\alpha} x^\alpha$ that connects the background frame $x^\mu =(t,\mathbf{x})$ to the instantaneous frame $x^{'\mu}=(t',\mathbf{x}')$ can be employed to determine what the accelerated observer measures. This method is clearly reasonable so long as the phenomena under consideration involve only pointlike coincidences involving classical point particles and electromagnetic rays that have, by definition, vanishing wavelengths.

Consider the reception of electromagnetic radiation by an accelerated observer. The incident wave packet consists, via Fourier analysis, of a spectrum of plane monochromatic waves each with propagation vector $k^\mu=(\omega ,\mathbf{k})$. The phase differential $d\varphi =k_\mu dx^\mu$ associated with each component of the wave packet is a Lorentz-invariant quantity; therefore, $k^{'\mu}=L^\mu _{\;\;\alpha}k^\alpha$. Thus the accelerated observer measures an \textit{instantaneous} spectrum with components $k^{'\mu}=(\omega ',\mathbf{k}')$ given by the standard formulas for the Doppler effect and aberration of starlight,
\begin{align}\label{eq:1} \omega '& = \gamma (\omega -\mathbf{v}\cdot \mathbf{k}),\\
\label{eq:2} \mathbf{k}'&=\mathbf{k}+\frac{\gamma -1}{v^2}(\mathbf{v}\cdot \mathbf{k})\mathbf{v}-\gamma \omega \mathbf{v},\end{align}
where $\mathbf{v}(t)$ is the instantaneous velocity of the observer and $\gamma$ is the corresponding Lorentz factor. To measure wave properties, the observer needs to register at least a few periods of the wave in order to make a reasonable determination. The $\mathbf{v}(t)$ in general changes from one instant to the next; therefore, one must conclude that phase invariance as well as Eqs.~\eqref{eq:1} and \eqref{eq:2} can be valid only in the geometric optics or eikonal \textit{limit} of wave motion corresponding to rays of radiation.

The standard treatment can be implemented in one of two equivalent ways. The first method involves, as already described above, making repeated Lorentz transformations to the instantaneous inertial rest frame of the observer. In the second method, one assigns a local tetrad frame to the accelerated observer. That is, instead of a continuous infinity of different inertial frames, a tetrad field is defined from the basis vectors of the inertial frames. The physical quantities measured by the accelerated observer---such as the electromagnetic field---are then the projections of various spacetime tensors on its tetrad frame. Let $\lambda^\mu_{\;\;(\alpha)}(\tau )$ be the orthonormal tetrad along the worldline of the accelerated observer. Here $\tau $ is the proper time, $\lambda^\mu_{\;\;(0)}=dx^\mu /d\tau$ is the unit timelike vector tangent to the observer's path, and $\lambda^\mu_{\;\;(i)}$, $i=1,2,3$, constitute the local spatial frame of the observer. According to the second method, the propagation vector measured by the accelerated observer is $k_{(\alpha)}=k_\mu \lambda^\mu_{\;\;(\alpha)}$.

The orthonormal tetrad frame of the fundamental inertial observers---i.e., those at rest---in the background global frame is given by $\bar{\lambda}^\mu_{\;\;(\alpha)}=\delta^\mu_{\;\;\alpha}$ and the tetrad frame of the accelerated observer at each instant $\tau$ is then
\begin{equation}\label{eq:3} \lambda_\mu^{\;\;(\alpha)} =L^\alpha_{\;\;\beta}\bar{\lambda}_\mu^{\;\;(\beta)} =L^\alpha_{\;\;\mu}.\end{equation}
Therefore,
\begin{equation}\label{eq:4} k^{(\alpha)}=k^\mu\lambda_\mu^{\;\;(\alpha)} =L^\alpha_{\;\;\mu}k^\mu=k^{'\alpha},\end{equation}
which illustrates the equivalence of the two methods in this case. In the rest of this paper, the second method will be employed.

To avoid unphysical situations, we generally assume that the acceleration of the observer starts at some initial instant $\tau_0$ and ends after a finite interval of time. Along the worldline of the observer, one can write
\begin{equation}\label{eq:5} \frac{d\lambda^\mu_{\;\;(\alpha)}}{d\tau}=\Phi _{(\alpha)} ^{\;\;\;\;\;(\beta)}\lambda^\mu_{\;\;(\beta)},\end{equation}
where $\Phi_{(\alpha )(\beta)}=-\Phi _{(\beta)(\alpha)}$ is the acceleration tensor. In fact, $\tilde{g}_{(i)}$, $\Phi _{(0)(i)}=\tilde{g}_{(i)}$, represent the observer's translational acceleration, while $\tilde{\Omega}_{(i)}$, $\Phi _{(i)(j)}=\epsilon_{(i)(j)(k)}\tilde{\Omega}^{(k)}$, represent the angular velocity of the observer's spatial frame with respect to a nonrotating (i.e., Fermi-Walker transported) tetrad frame along the worldline. The spacetime invariants $\tilde{\mathbf{g}}$ and $\tilde{\boldsymbol{\Omega}}$ represent the rate of variation of the observer's local tetrad frame; hence, an accelerated observer has certain intrinsic length $(\mathcal{L})$ and time $(\mathcal{L}/c)$ scales associated with its motion that can be constructed from $\tilde{\mathbf{g}}$ and $\tilde{\boldsymbol{\Omega}}$~\cite{7}. For instance, for a static observer on the Earth, $\mathcal{L}=c^2/g_\oplus\cong 1$ light year is its translational acceleration length and $\mathcal{L}=c/\Omega_\oplus\cong 28$ A.U. is its rotational acceleration length. Nonlocal special relativity is based on the assertion that the locality postulate is valid in the limit of vanishing $\lambda /\mathcal{L}$, where $\lambda$ is the characteristic wavelength of the phenomenon under observation; in practice, $\lambda /\mathcal{L}$ is generally very small and deviations from locality are then expected to be proportional to $\lambda /\mathcal{L}$.

Inertial observers are naturally endowed with tetrad frames and the designation of tetrad frames for accelerated observers follows from the hypothesis of locality. The actual establishment of a local reference frame by experiment is altogether a different matter. Studies of length and time measurements by accelerated observers indicate the existence of limitations associated with the notions of \textit{standard clock} and \textit{rod}---i.e., devices that function exactly in accordance with the postulate of locality~\cite{7}. We therefore \textit{define} the proper time $\tau$ of an accelerated observer along its worldline via
\begin{equation}\label{eq:6} \tau =\int(1-v^2)^{1/2}dt,\end{equation}
which can in principle be determined by the background fundamental inertial observers along the worldline; similarly, we \textit{define} $\lambda^\mu_{\;\;(\alpha)}$ via the integration of Eq.~\eqref{eq:5} with initial data given at $\tau _0$. In this way, the accelerated observer is \textit{assigned} a tetrad frame in nonlocal special relativity.

The standard treatment can be extended to curvilinear coordinate systems in Minkowski spacetime by means of tensor calculus; in fact, no new physical assumption is needed for this purpose. A further natural extension to general relativity is ensured by Einstein's local principle of equivalence~\cite{7}.

\section{Locality: Improved Approach}\label{sec:3}

In the revised and improved treatment of the reception of electromagnetic radiation, the locality postulate is applied directly to the field, rather than the propagation vector. This involves the projection of the field on the tetrad frames of the accelerated observers, namely,
\begin{equation}\label{eq:7}
F_{(\alpha)(\beta)} =F_{\mu\nu}\lambda^\mu_{\;\;(\alpha)} \lambda^\nu_{\;\;(\beta)}.\end{equation}
Then, $F_{(\alpha )(\beta)}$ is Fourier analyzed in order to determine its contents in terms of frequency and wave vector of the radiation~\cite{8}. For inertial observers, the standard and revised treatments give the same answer, but the situation is different for accelerated observers. We note that the revised treatment contains an element of nonlocality as it involves Fourier analysis, which is a nonlocal process. Consequently, the improved treatment goes beyond the eikonal limit of vanishing wavelength.

The main implication of the new approach for observers that rotate uniformly with frequency $\Omega >0$ is that the measured frequency is now
\begin{equation}\label{eq:8} \omega '_M=\gamma (\omega -M\Omega),\quad M=0,\pm 1,\pm 2,\dots .\end{equation}
Here $\hbar M$ is the total (orbital plus spin) angular momentum of the radiation along the axis of rotation of the observer~\cite{9}. The general coupling of rotation with the angular momentum of the radiation field has been the subject of a number of basic investigations; in particular, the difference between Eq. (8) and the standard Doppler
formula (1), known as the angular Doppler shift in classical optics, has
been studied in connection with the frequency shift resulting from the
passage of polarized radiation through a rotating spin flipper, the
spin-orbit interaction, the Berry phase, and the spin Hall effect---see~\cite{10} and the references therein. In the eikonal approximation, Eq.~\eqref{eq:8} may be written as $\omega '=\gamma (\omega -\mathbf{J}\cdot \boldsymbol{\Omega})$ with $\mathbf{J}=\mathbf{r}\times \mathbf{k} +s \hat{\mathbf{H}}$, where $\hbar \mathbf{J}$ is the total angular momentum vector, $s$ is the spin, and $\hat{\mathbf{H}} =\pm \hat{\mathbf{k}}$ is the helicity vector of the photon. Thus Eq.~\eqref{eq:8} in the eikonal approximation reduces to the Doppler effect together with the term $-\gamma s\hat{\mathbf{H}}\cdot \boldsymbol{\Omega}$, which indicates helicity-rotation coupling---see Eq.~\eqref{eq:10} below. This phenomenon has been discussed in detail elsewhere~\cite{9,11}. It has extensive observational support; moreover, its existence implies that the phase of the radiation is not a Lorentz-invariant quantity in the revised treatment.

Equation~\eqref{eq:8} has two unusual aspects: for $\omega \leq M\Omega$, $\omega '$ can be negative or zero. A negative frequency according to the rotating observers poses no difficulty and is simply a consequence of the circumstance that the Hamiltonian in this case is not bounded from below. A thought experiment was presented in~\cite{9} to show that a negative $\omega '$ is consistent with the fact that the temporal order of events should be independent of the choice of observer.

The second special feature of Eq.~\eqref{eq:8} is that
\begin{equation}\label{eq:9} \omega '_M=0 \text{ for } \omega =M\Omega,\quad M\neq 0.\end{equation}
That is, by a mere rotation of frequency $\Omega =\omega /M $, $M\neq 0$, the rotating observer can stand completely still with the incident radiation. The situation here is quite similar to the pre-relativistic Doppler formula, where an observer could be comoving with light. This influenced Einstein's approach to relativity, as mentioned in his autobiographical notes---see page 53 of~\cite{12}. To avoid this fundamental difficulty, it is important to formulate the theory of accelerated observers in such a way that an accelerated observer cannot stay completely at rest with an electromagnetic wave. We return to this point at the end of section~\ref{sec:4}.

In the eikonal approximation $(\omega \gg \Omega)$, the modified expressions for Doppler effect and aberration have been derived in~\cite{8,9,11} and the results are
\begin{align} \label{eq:10} \omega '&=\gamma [(\omega -s\hat{\mathbf{H}}\cdot \boldsymbol{\Omega})-\mathbf{v}\cdot \mathbf{k}],\\
\label{eq:11}\mathbf{k}'&=\mathbf{k}+\frac{\gamma -1}{v^2} (\mathbf{v}\cdot \mathbf{k})\mathbf{v}-\gamma (\omega -s\hat{\mathbf{H}} \cdot \boldsymbol{\Omega} )\mathbf{v},\end{align}
where $s=1$ for the photon and $s=2$ for the graviton~\cite{13}. To illustrate the new terms in Eqs.~\eqref{eq:10} and \eqref{eq:11}, let us first note that the standard formulas are recovered for $\mathbf{k}\cdot \boldsymbol{\Omega}=0$. Hence, we consider a simple situation involving normal incidence with $\mathbf{k}=\omega \hat{\boldsymbol{\Omega}}$, so that
\begin{equation}\label{eq:12} \omega '=\gamma (\omega \mp s\Omega),\quad \mathbf{k}'-\mathbf{k}=-\gamma (\omega \mp s\Omega)\mathbf{v}.\end{equation}
The expression for frequency in Eq.~\eqref{eq:12} happens to be exact in this case and the spin-rotation coupling part has been verified for $\omega \gg \Omega$ in the GPS, where it accounts for the phenomenon of phase wrap-up~\cite{14}. For the aberration part of Eq.~\eqref{eq:12}, we note that with respect to the direction of incidence of the wave, $\mathbf{k}'_{\parallel}=\mathbf{k}$ and $    \mathbf{k}'_{\bot} =-\gamma (\omega \mp s\Omega )\mathbf{v}$; hence the aberration angle $\alpha =\tan^{-1}(|\mathbf{k}'_\bot |/|\mathbf{k}'_{\parallel}|)$ is in this case $\alpha_\pm =\tan^{-1}[\gamma v(1\mp s\Omega /\omega )]$. The deviations from the standard results in Eq.~\eqref{eq:12} are proportional to $\Omega /\omega =\lambdabar /\mathcal{L}$, as expected. Further discussions of helicity-rotation coupling are contained in Ref.~\cite{15}. We now turn to the genesis of acceleration-induced nonlocality.

\section{Bohr-Rosenfeld Principle}\label{sec:4}

Bohr and Rosenfeld demonstrated that the formalism of quantum electrodynamics is fully compatible with the quantum theory of measurement~\cite{16}. Their starting point was the important observation---which we refer to as the ``Bohr-Rosenfeld principle"---that although in \textit{classical} electrodynamics one deals with the field $F_{\mu\nu}(x)$ defined at an event $x$ in spacetime, such an idealization is devoid of \textit{immediate} physical significance; indeed, only averages of such field components over finite spacetime regions $\Delta$, that is,
\begin{equation}\label{eq:13} \langle F_{\mu\nu}\rangle =\frac{1}{\Delta} \int_{\Delta} F_{\mu\nu} (x)d^4x,\end{equation}
can be physically meaningful~\cite{16}. Bohr and Rosenfeld reached this conclusion following a critical analysis of how the electromagnetic field is in fact measured in classical physics via the Lorentz force law~\cite{16}.

It is important to observe that in the Bohr-Rosenfeld argument in particular, and in quantum measurement theory in general, the nature of the observer is not explicitly specified, since it is always implicitly assumed from the outset that all physical experiments are performed by \textit{inertial} observers. The Bohr-Rosenfeld principle, which we take to be valid for \textit{all} observers, is consistent with the electrodynamics of inertial systems, because an inertial observer has no basic \textit{classical} length or time scales, in sharp contrast with the intrinsic scales of an accelerated observer. To illustrate this point, imagine, for instance, the reception of an incident electromagnetic wave by an accelerated observer. A few periods of the wave must be registered by the observer before any approximate determination of wave properties---such as amplitude, frequency, wave vector, and polarization---even becomes possible. During this process, the local tetrad frame of the noninertial observer in general varies continuously. What is the electromagnetic field $\mathcal{F}_{(\alpha )(\beta )}(\tau )$ that is measured by the accelerated observer?

To answer this question, let us first recall that in special relativity the issue is treated via the locality postulate, which implies that $\mathcal{F}_{(\alpha )(\beta)}=F_{(\alpha)(\beta)}$ at each instant $\tau $ along the worldline. This \textit{pointwise} field measurement contradicts the Bohr-Rosenfeld principle. To resolve this basic conflict, we consider instead the most general linear connection between the accelerated observer and the sequence of comoving inertial observers. The general linear relation between $\mathcal{F}_{(\alpha)(\beta)}$ and $F_{(\alpha)(\beta)}$ that is consistent with causality can be expressed as
\begin{equation}\label{eq:14} \mathcal{F}_{(\alpha )(\beta)}(\tau )=F_{(\alpha )(\beta)}(\tau )+u(\tau -\tau _0) \int^\tau _{\tau _0} \mathcal{K}_{(\alpha)(\beta)}^{\;\;\;\;\;\;\;\;\;(\gamma )(\delta)} (\tau ,\tau ')F_{(\gamma)(\delta)} (\tau ')d\tau '.\end{equation}
Here $u(t)$ is the unit step function such that $u(t)=1$ for $t>0$ and $u(t)=0$ for $t<0$. Moreover, the kernel is assumed to be directly dependent upon the acceleration of the observer. A simple estimate then suggests that the nonlocal term in Eq.~\eqref{eq:14} is proportional to $\lambda /\mathcal{L}$, so that locality is recovered for $\lambda /\mathcal{L}\to 0$. The nonlocal part in ansatz~\eqref{eq:14} is essentially an average---in the spirit of the Bohr-Rosenfeld principle---over the past worldline of the observer such that the weight function is directly proportional to acceleration.

The main criterion employed in nonlocal special relativity to determine the kernel is that an accelerated observer can never be comoving with an electromagnetic wave---see~\cite{3} and the references cited therein for a more complete discussion. In particular, the situation expressed in Eq.~\eqref{eq:9}, which follows from the revised approach, should be forbidden in the nonlocal theory. These issues are discussed in the following section, where we consider the reception of radiation by uniformly rotating observers in accordance with nonlocal special relativity. The derivation of nonlocal effects presented below is general, in contrast
to previous work that was restricted to the particular case of normal
incidence.

\section{Nonlocality: Uniform Rotation}\label{sec:5}

Imagine an observer in the $(x,y)$ plane following a straight line $x=r>0$ with constant speed $v$ for $-\infty <t<0$ such that $y=vt$ and at $t=0$ it is forced to move on a circle of radius $r$ in the positive sense with frequency $\Omega =v/r$. We are interested in the motion of the observer for $t\geq 0$, when its natural tetrad frame with respect to the global inertial coordinates $(t,x,y,z)$ is given by
\begin{align}\label{eq:15} \lambda^\mu_{\;\;(0)}&=\gamma (1,-v\sin \phi ,v\cos \phi ,0),\\
\label{eq:16} \lambda^\mu_{\;\;(1)}&= (0,\cos \phi ,\sin \phi ,0),\\
\label{eq:17} \lambda^\mu_{\;\;(2)}&=\gamma (v,-\sin \phi ,\cos \phi ,0),\\
\label{eq:18}\lambda^\mu_{\;\;(3)}&=(0,0,0,1).\end{align}
Here $\phi =\Omega t=\gamma \Omega \tau$, $\gamma$ is the observer's Lorentz factor, and $\tau$ is its proper time such that $\tau =0$ at $t=0$. Thus $\lambda^\mu_{\;\;(0)}$ is the $4$-velocity vector of the observer and its spatial frame $\lambda^\mu_{\;\;(i)}$, $i=1,2,3$, is given by the radial, tangential, and normal unit vectors, respectively. The purpose of this section is to work out the consequences of nonlocal special relativity for the description of incident electromagnetic radiation by such uniformly rotating observers.

The incident radiation field in the background inertial frame is characterized, for simplicity, by its vector potential $A^\alpha =(0,\mathbf{A})$ in the Coulomb gauge $(\boldsymbol{\nabla}\cdot \mathbf{A}=0)$. In terms of electromagnetic modes of definite momentum and helicity, $\mathbf{A}$ may be expressed as
\begin{equation}\label{eq:19} \mathbf{A}(t,\mathbf{r})=\sum_{\mathbf{k}\;\;\epsilon} \left( \frac{2\pi \hbar}{\omega V}\right)^{1/2} (a_{\mathbf{k}\epsilon}\mathbf{q}_{\mathbf{k}\epsilon} +a^\dagger_{\mathbf{k}\epsilon }\mathbf{q}^\ast _{\mathbf{k}\epsilon} ),\end{equation}
where $a_{\mathbf{k}\epsilon}$ and $a^\dagger_{\mathbf{k}\epsilon}$ are the photon annihilation and creation operators, respectively, and
\begin{equation}\label{eq:20} \mathbf{q}_{\mathbf{k}\epsilon}=\mathbf{e}_{\mathbf{k}\epsilon} e^{-i\omega t+i\mathbf{k}\cdot \mathbf{r}}.\end{equation}
Here $\omega=k$, $\epsilon =\pm 1$ represents the helicity of the photon, and $V$ is the volume of space within a large cube; moreover, $\mathbf{e}_{\mathbf{k}\epsilon}$ is the unit circular polarization basis for a photon of wave vector $\mathbf{k}$ such that $\mathbf{k}\cdot \mathbf{e}_{\mathbf{k}\epsilon} =0$. All of the operations below involving the electromagnetic field will be linear and the only quantity in Eq.~\eqref{eq:19} that will be affected in the course of our calculations will be $\mathbf{q}_{\mathbf{k}\epsilon}$; therefore, to simplify matters without any loss in the generality of our final conclusions, we will focus attention on a single wave vector $\mathbf{k}$. Furthermore, in order to express complicated expressions in a simple form, we introduce for the sum in Eq.~\eqref{eq:19} involving the chosen $\mathbf{k}$ the notation
\begin{equation}\label{eq:21} \mathbf{A}(t,\mathbf{r})=\mathcal{S}\{\mathbf{q}_{\mathbf{k}\epsilon}\},\end{equation}
which will be used below.

To illustrate the difference between local and nonlocal predictions, we first calculate the field according to the sequence of momentarily comoving inertial observers along the worldline $(\tau \geq 0)$,
\begin{equation}\label{eq:22} A_{(\alpha)} (\tau )=A_\mu\lambda^\mu_{\;\;(\alpha)}.\end{equation}
This involves the projection of the incident field, evaluated at the position of the rotating observer, onto its tetrad frame; moreover, we use the formula
\begin{equation}\label{eq:23} e^{i\mathbf{k}\cdot \mathbf{r}}=4\pi \sum_{\ell m} i^\ell j_\ell (kr)Y_{\ell m}^\ast (\hat{\mathbf{k}} )Y_{\ell m} (\hat{\mathbf{r}}),\end{equation}
where $j_\ell $, $\ell =0,1,2,\dots $, are spherical Bessel functions. Next, we determine the field according to the rotating observer via the nonlocal theory $(\tau \geq 0)$,
\begin{equation}\label{eq:24} \mathcal{A}_{(\alpha)} (\tau )=A_{(\alpha)} (\tau )+\int^\tau _0\mathcal{K}_{(\alpha)}^{\;\;\;\;\;(\beta )} (\tau ,\tau ') A_{(\beta)} (\tau ')d\tau '.\end{equation}
This is the analog of Eq.~\eqref{eq:14} for the vector potential with $\tau _0=0$. In this case, we adopt the kernel
\begin{equation}\label{eq:25} \mathcal{K}_{(\alpha)}^{\;\;\;\;\;(\beta)} (\tau ,\tau ')=-\Phi _{(\alpha)}^{\;\;\;\;\;(\beta)}(\tau '),\end{equation}
which ensures that a radiation field never stands completely still with respect to an observer (see~\cite{3} and references therein). We find from Eq.~\eqref{eq:5} and the tetrad frame \eqref{eq:15}-\eqref{eq:18} that the only nonzero components of the acceleration tensor are given by
\begin{align}\label{eq:26}\Phi_{(0)(1)}&=-\Phi _{(1)(0)} =-\beta \gamma^2\Omega ,\\
\label{eq:27}\Phi_{(1)(2)}&=-\Phi_{(2)(1)}=\gamma^2\Omega.\end{align}

According to the comoving inertial observers
\begin{align}\label{eq:28} A_{(0)}&=vA_{(2)},\quad A_{(1)}=\cos \phi A_x+\sin \phi A_y,\\
\label{eq:29} A_{(2)} &=\gamma (-\sin \phi A_x +\cos \phi A_y),\quad A_{(3)}=A_z,\end{align}
where $A_x$, $A_y$, and $A_z$ are the Cartesian components of the vector potential evaluated at the position of the rotating observer.

To work out the explicit expressions for $A_{(\alpha)}$ and $\mathcal{A}_{(\alpha)}$, it is helpful to choose the Cartesian coordinate system such that $\hat{\mathbf{k}}=(\vartheta ,0)$ in spherical polar coordinates. Then, $\hat{\mathbf{k}}$, $\hat{\mathbf{n}}$, and $\hat{\mathbf{y}}$ form an orthonormal triad, where $\hat{\mathbf{n}}=\cos\vartheta \hat{\mathbf{x}} -\sin \vartheta \hat{\mathbf{z}}$. The circular polarization states are given by $(\hat{\mathbf{n}}\pm i\hat{\mathbf{y}} )/\sqrt{2}$, or
\begin{equation}\label{eq:30} \mathbf{e}_{\mathbf{k}\epsilon} =\frac{1}{\sqrt{2}} (\cos \vartheta \hat{\mathbf{x}} +\epsilon i\hat{\mathbf{y}}-\sin \vartheta \hat{\mathbf{z}}).\end{equation}
It follows that
\begin{align}\label{eq:31} A_{(1)}&=\mathcal{S}\{ \zeta_\epsilon ^+ Q^+_{\mathbf{k}\epsilon} +\zeta ^-_{\epsilon} Q^-_{\mathbf{k}\epsilon} \},\\
\label{eq:32}A_{(2)}&=i\gamma \mathcal{S} \{ \zeta ^+_\epsilon Q^+_{\mathbf{k}\epsilon }-\zeta^-_\epsilon Q^-_{\mathbf{k}\epsilon }\},\\
\label{eq:33} A_{(3)}&=-\sin \vartheta \mathcal{S}\{ Q^0_{\mathbf{k}\epsilon}\}.\end{align}
Here
\begin{align}\label{eq:34}\zeta^\pm_\epsilon &=\frac{1}{2} (\pm \epsilon +\cos \vartheta),\\
\label{eq:35} Q^\sigma_{\mathbf{k}\epsilon} &=\frac{4\pi}{\sqrt{2}}\sum_{\ell m}i^\ell j_\ell (kr)Y^\ast_{\ell m} (\vartheta ,0)Y_{\ell m}\left( \frac{\pi}{2},0\right)e^{-i\omega '_{m\sigma}\tau }.\end{align}
Moreover, we have introduced the spin parameter $\sigma =0$, $\pm 1$, and
\begin{equation}\label{eq:36} \omega '_{m\sigma}=\gamma [\omega -(m+\sigma )\Omega ],\end{equation}
which is the same as Eq.~\eqref{eq:8} with $M=m+\sigma$, where the contributions of orbital $(m)$ and spin $(\sigma)$ angular momentum of the photon (in units of $\hbar$) along the axis of rotation of the observer have been made explicit. To simplify notation, $\sigma$ is also expressed as plus, minus, or zero.

The result of the nonlocal theory, given by Eqs.~\eqref{eq:24}-\eqref{eq:27}, can be similarly expressed as $\mathcal{A}_{(0)}=v\mathcal{A}_{(2)}$,
\begin{align}\label{eq:37} \mathcal{A}_{(1)}&=\mathcal{S} \{ \zeta ^+_\epsilon W^+_{\mathbf{k}\epsilon} +\zeta^-_\epsilon W^-_{\mathbf{k}\epsilon}\},\\
\label{eq:38} \mathcal{A}_{(2)}&=i\gamma \mathcal{S}\{ \zeta^+_\epsilon W^+_{\mathbf{k}\epsilon}-\zeta_\epsilon^-W^-_{\mathbf{k}\epsilon}\},\\
\label{eq:39} \mathcal{A}_{(3)}&=-\sin \vartheta \mathcal{S}\{W^0_{\mathbf{k}\epsilon}\},\end{align}
where
\begin{equation}\label{eq:40} W^\sigma_{\mathbf{k}\epsilon} =\frac{4\pi}{\sqrt{2}} \sum_{\ell m} i^\ell j_\ell (kr)Y^\ast _{\ell m} (\vartheta ,0)Y_{\ell m}\left( \frac{\pi}{2},0\right) f_{m\sigma} (\tau )\end{equation}
and
\begin{equation}\label{eq:41} f_{m\sigma}(\tau )=\frac{\omega '_{m0}}{\omega '_{m\sigma}}e^{-i\omega '_{m\sigma}\tau }-\frac{\sigma \gamma \Omega}{\omega '_{m\sigma}}.\end{equation}
In Eqs.~\eqref{eq:35} and \eqref{eq:40}, $Y_{\ell m}\left( \frac{\pi}{2},0\right)$ vanishes when $\ell +m$ is odd. Otherwise, for $\ell +m=2n$, $n=0,1,2,\dots$,
\begin{equation}\label{eq:42} Y_{\ell m}\left( \frac{\pi}{2},0\right) =(-1)^n \left( \frac{2\ell +1}{4\pi}\right)^{1/2} \frac{[(2n)! (\ell -m)!]^{1/2}}{(2n)!!(\ell -m)!!}.\end{equation}
Moreover, for normal incidence $Y_{\ell m}(\pm \hat{\mathbf{z}})=0$, unless $m=0$.

The difference between the local and nonlocal results is simply that $f_{m\sigma}(\tau )$ appears in Eq.~\eqref{eq:40} in place of $\exp (-i\omega '_{m\sigma}\tau )$ in Eq.~\eqref{eq:35}. These functions are indeed different only when $\tau >0$ and $\sigma =\pm 1$; that is, $Q_{\mathbf{k}\epsilon}^0=W^0_{\mathbf{k}\epsilon}$. In case $\omega '_{m\sigma}=0$, the corresponding term in Eq.~\eqref{eq:35} loses its temporal dependence; however, for $\sigma =\pm 1$, $f_{m\sigma}$ exhibits resonance behavior, namely,
\begin{equation}\label{eq:43} f_{m\sigma}(\tau )\to 1-i\sigma \gamma \Omega \tau \quad \text{as}\quad \omega '_{m\sigma} \to 0.\end{equation}

For orbital angular momentum parameters $(\ell ,m)$, the measured frequency for each $m$ splits, due to photon's spin, into three components. It follows from Eq.~\eqref{eq:41} that nonlocality produces a change in the wave amplitude for $\sigma =\pm 1$ given by the factor $\omega '_{m0}/\omega '_{m\sigma}$. We denote this ratio by $\rho$,
\begin{equation}\label{eq:44} \rho =\frac{\omega -m\Omega}{\omega -(m+\sigma)\Omega },\end{equation}
which differs from unit only when $\sigma =\pm 1$. Away from resonance with $\omega '_{m\sigma}>0$ (including $\omega '_{m0}>0$), $\rho >1$ for $\sigma =1$ and $\rho <1$ for $\sigma =-1$. Thus, as a result of nonlocality, the amplitude increases (decreases) if the component of photon spin along the axis of rotation of the observer is positive (negative). This circumstance reverses for $\omega '_{m\sigma}<0$; that is the amplitude decreases (increases) if the spin component along the rotation axis is positive (negative).

In the special case of normal incidence, where $m=0$ and $\sigma =\pm 1$, the whole treatment is considerably simplified. In fact, this case has been extensively studied before; in particular, Bohr's correspondence principle has been employed to demonstrate that quantum mechanics in the limit of large quantum numbers is qualitatively consistent with our nonlocal considerations in connection with Eqs.~\eqref{eq:43} and \eqref{eq:44} for normal incidence~\cite{17}. The case of oblique incidence treated in this paper constitutes a generalization of previous work---see~\cite{17} and references therein.

In view of Eqs.~\eqref{eq:8}, \eqref{eq:9}, and \eqref{eq:36}, it is interesting to consider briefly photon states of definite total angular momentum $\hbar J$, where $J=1,2,3,\dots$. In this case, the orbital angular momentum $\hbar \ell $ is such that $\ell =J+1$, $J$, or $J-1$. The vector potential of the radiation field can be expressed in terms of these modes as
\begin{equation}\label{eq:45} \mathbf{A}=\sum_{\omega JM\delta} (a^\delta_{\omega JM}\mathbf{A}^\delta_{\omega JM}+a^{\delta \dagger}_{\omega JM}\mathbf{A}^{\delta \ast}_{\omega JM}),\end{equation}
where $\delta$ stands for either electric multipole radiation of parity $(-1)^{J+1}$ or magnetic multipole radiation of parity $(-1)^J$. The precise expression for each $\mathbf{A}^\delta_{\omega JM}$ in terms of vector spherical harmonics is given in section~135 of Ref.~\cite{18}. The function $\mathbf{A}^\delta_{\omega JM}$ depends upon time as $\exp (-i\omega t)$; therefore, when $\mathbf{A}^\delta_{\omega JM}$ is evaluated at the position of the rotating observer, where $\phi =\gamma \Omega \tau$, its temporal dependence becomes $\exp (-i\omega '_M\tau )$. It follows from Eqs.~\eqref{eq:28} and \eqref{eq:29} that at the position of the observer, $A_{(1)}=\mathbf{A}\cdot \hat{\mathbf{r}}$, $A_{(2)}=\gamma \mathbf{A}\cdot \hat{\boldsymbol{\phi}}$, and $A_{(3)}=-\mathbf{A}\cdot \hat{\boldsymbol{\theta}}$, so that for a given mode with $\omega$, $J$, $M$, and $\delta$, the corresponding $A_{(\alpha)}$ varies with time as $\exp (-i\omega '_M\tau )$. Hence $A_{(\alpha)}$ for this mode becomes constant in time for $\omega '_M=0$, but the corresponding $\mathcal{A}_{(\alpha)}$ varies linearly with proper time in this case as a consequence of Eqs.~\eqref{eq:24}-\eqref{eq:27}. Thus the incident photon and the rotating observer cannot be at rest with respect to each other. It follows that in nonlocal special relativity, Eq.~\eqref{eq:8} is valid except when $\omega =M\Omega$, in which case the reception of radiation is dominated by resonance.

\section{Discussion}\label{sec:6}

Nonlocality excludes the possibility that an accelerated observer can stay completely at rest with an incident electromagnetic wave; for a uniformly rotating observer, the resonance behavior of the field has been studied in general for $\omega '_M\to 0$. Furthermore, the wave amplitude is affected by nonlocality. For a rotating observer with measured frequency $\omega '_M>0$, the amplitude is higher (lower) if the component of photon spin along the axis of rotation of the observer $(\hbar \sigma)$ is positive (negative); that is, if it is in the same (opposite) sense as the rotation of the observer.

There is at present a woeful lack of reliable observational data regarding Maxwell's theory in accelerated frames of reference. Hopefully, the results of this paper as well as Ref.~\cite{3} will provide a much-needed boost to experimental studies of the electrodynamics of accelerated systems.

\section*{Acknowledgement}

I am grateful to F.W. Hehl for helpful discussions.

\end{document}